\documentclass[%
 aip,
 amsmath,amssymb,
 reprint,%
]{revtex4-1}

\usepackage{graphicx}
\usepackage{dcolumn}
\usepackage{bm}

\usepackage[utf8]{inputenc}
\usepackage[T1]{fontenc}
\usepackage{mathptmx}
\usepackage{etoolbox}

\makeatletter
\def\@email#1#2{%
 \endgroup
 \patchcmd{\titleblock@produce}
  {\frontmatter@RRAPformat}
  {\frontmatter@RRAPformat{\produce@RRAP{*#1\href{mailto:#2}{#2}}}\frontmatter@RRAPformat}
  {}{}
}%
\makeatother
\begin{document}

\preprint{AIP/123-QED}

\title[Gravity-related collapse]{Gravity-related collapse of the wave function and spontaneous heating: revisiting the experimental bounds}
\author{Andrea Vinante}
\affiliation{ 
CNR - Istituto di Fotonica e Nanotecnologie and Fondazione Bruno Kessler, I-38123 Povo, Trento, Italy
}
\email{anvinante@fbk.eu}

\author{Hendrik Ulbricht}
\affiliation{School of Physics and Astronomy, University of Southampton, SO17 1BJ, Southampton, UK}
\email{h.ulbricht@soton.ac.uk}

\date{\today}

\begin{abstract}
The possibility that the collapse of the wave function in quantum mechanics is a real and ultimately connected to (classical) gravity has been debated for decades, with main contributions by Di\'osi and Penrose. In particular, Di\'osi proposed a noise-based dynamical reduction model, which captures the same orders of magnitude for the collapse time suggested by Penrose based on heuristic arguments. This is known in literature as the DP model (Di\'osi-Penrose). A peculiarity of the DP model is the prediction of spontaneous heating of matter, which can be tested without the need for massive quantum superpositions. Notably, a very similar effect is predicted by recent theoretical approaches to gravity as a classical-only information channel. Here, we reconsider the current constraints on the DP model from spontaneous heating, by analyzing experimental situations not properly considered before. We argue that the the parameter-free version of the DP model is close to be ruled out by standard heat leak measurements at ultralow temperature, with a conclusive exclusion likely within reach with existing technology. This result would strengthen a recent claim of exclusion inferred by spontaneous x-ray emission experiments, which relies on the somewhat stronger assumption that the DP noise field is white up to x-ray frequencies. 
\end{abstract}

\maketitle

%

The universality of quantum mechanics, and in particular the fact that the quantum superposition principle holds up to arbitrarily large scales, is often given for granted in modern physics. Several notable scientists have challenged this view. In particular, Roger Penrose has argued that the quantum superposition principle is in tension with the covariance principle of General Relativity, suggesting that the former may need to be dropped, rather than assuming that gravity has to be quantized \cite{penrose1,penrose2}. Quantum superpositions of massive objects in distinct locations create a quantum superposition of spacetimes with a difference in gravitational energy $\Delta E$. Penrose suggests that such superposition will decay on a time scale of the order of $\tau=\hbar/\Delta E$, where $\hbar$ is the Planck constant. However, he did not come up with a precise dynamical model for this mechanism, despite early attempts of connections to a Schr\"odinger-Newton dynamics \cite{penrose2}.

A similar result was obtained by Di\'osi using different arguments, following the basic scheme of collapse models \cite{DP1,DP2}. Initially developed by Ghirardi and collaborators \cite{GRW,CSL}, collapse models postulate the existence of a universal stochastic noise that modifies the standard deterministic dynamics of a quantum system. The nonlinear coupling to matter is tailored so as to produce the stochastic localization of the wave function (i.e. a collapse in position), in agreement with the Born rule. On a general level, collapse models are one of the possible ways to solve the problem of macro-objectification in quantum mechanics, sometimes referred to as the measurement problem. Assuming that the stochastic noise is related to gravity, and with further heuristic arguments, Di\'osi proposed a dynamical stochastic model which is found to reproduce Penrose's prediction for the decay time of a massive superposition. In literature, this is usually referred to as the Di\'osi-Penrose (DP) model \cite{collapsereview2}.

While highly suggestive, the DP model suffers from an annoying side effect, that is in common with other collapse models. The stochastic noise introduced {\it ad-hoc} in the quantum dynamics leads to a universal heating, which shows up at different levels: as a universal force noise acting on mechanical systems \cite{collett, nimmrichter, diosi2016, vinante1, vinante2, vinante3}, as a spontaneous heating of matter \cite{adlervinante}, or as spontaneous radiation from noise-driven charged particles \cite{adlerX}. For the specific case of the DP model, the rate of increase of thermal energy for a system of mass $m$ can be calculated explicitly as \cite{GGR,diosi2014}:
\begin{equation}    \label{DPheating}
    \frac{dE}{dt}=\frac{G \hbar m}{4 \sqrt{\pi} R_0^3}
\end{equation}
where $G$ is the gravitational constant and $R_0$ is a renormalization length which needs to be introduced in order to avoid divergences.
Fig. 1 shows the dependence of the specific power per unit mass on $R_0$, in a range relevant to this paper.
\begin{figure}[!ht]
\includegraphics[width=8.6cm]{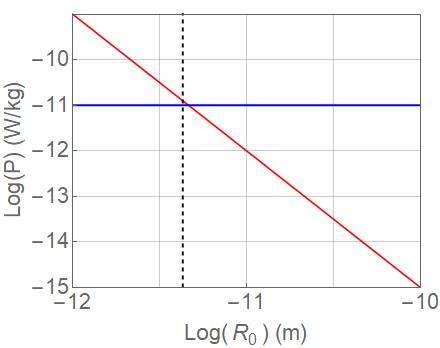}
\caption{Red line: specific power per unit mass predicted by the DP model as a function of the renormalization length $R_0$, in the range relevant to this paper. Blue horizontal line: upper limit on time-independent unmodeled specific power inferred from heat leak measurements at ultralow temperature. Black dashed line: nuclear wave function spread in copper at low temperature. This parameter sets the length scale of the parameter-free DP model in the experiment.}  \label{decay}
\end{figure}

$R_0$ was introduced by Ghirardi {\it et al} \cite{GGR} as a fundamental free parameter of the DP model. For concreteness, they proposed the value $R_0=10^{-7}$ m, identical to the characteristic length of other collapse models. With this choice, the heating from the DP model becomes exceedingly small, far from being experimentally relevant, at the price of introducing a free parameter. However, one may still think at the DP model as being parameter-free, by defining $R_0$ as a proper length, dependent on the system and characterizing the mass density spatial resolution. An early choice by Di\'osi was indeed to take as $R_0$ the nuclear size, i.e. $R_0 \approx 10^{-14}$ m. Following an argument suggested by Penrose, when considering gravitational effects of a quantum system one should instead take the mass density as $\rho(\mathbf{r,t})=m \psi (\bold{r},t)$ where $\psi ({\bold{r},t})$ is the wave function \cite{penrose2014} as function of position $\bold{r} $ and time $t$. Accordingly, the actual mass resolution in solids would be set by the thermal amplitude (or quantum amplitude, at low temperature) of the nuclear wave function in the quasiharmonic trap set by the crystal structure. This is of the order of $R_0\approx 10^{-12}$ m for typical solid matter.

The experimental bounds on the DP model reported in literature rule out the earlier Di\'osi guess $R_0\approx 10^{-14}$ m completely. For instance, the ultralow force noise measured in the space mission LISA Pathfinder \cite{helou} sets $R_0>8.2 \times 10^{-14}$ m, using the improved noise values reported in Ref. \citenum{LISA2}. However, very recently even the more relaxed value of $R_0$ based on the Penrose argument was claimed to be ruled out by Donadi {\it et al} \cite{donadi}, based on measuring very low levels of x-ray spontaneous emission in ultrapure Germanium samples \cite{donadi}. Here, we point out that the bound derived in Ref. \citenum{donadi} is based on the explicit assumption that the stochastic noise is strictly Poissonian, i.e., with white noise spectrum up to $10^{-18}$ Hz. Such an assumption can be easily evaded by postulating that the stochastic noise has a high frequency cutoff. This argument has been often supported, among others by Adler, in the context of the CSL model \cite{adler}. According to Adler, experimental bounds set by low frequency experiments are therefore more relevant. In this sense, in order to provide a conclusive and convincing exclusion, it is crucial to test the DP model at different frequencies, spanning over many orders of magnitude.

In the following, we will show that the parameter-free DP model with the Penrose's choice for $R_0$ is close to be ruled out as well by the most accurate measurements of heating effects in solid matter available in laboratory. This reduces the probed characteristic frequency from $10^{-18}$ Hz, typical of x-ray photons, down to $10^{-12}$ Hz, typical of solid matter phonons \cite{adlervinante}.

We start by observing that Eq.~(\ref{DPheating}), which can be easily derived for a gas of noninteracting particles \cite{smirne2014}, is valid as well in a solid state system. In fact, Di\'osi has derived the same equation by modeling an elastic body as a collection of noninteracting acoustic modes \cite{diosi2014}.

Next, let us investigate the best candidate experiments to probe the smallest possible excess heating. Tilloy and Stacey \cite{tilloy} have derived a bound from the equilibrium temperature of the coldest observable neutron stars. This corresponds to a radiated power per unit mass $P \approx 100$ nW/kg, which leads, through Eq.~(\ref{DPheating}), to a lower bound $R_0 \gtrsim 10^{-13}$ m. The latter is reported in the recent paper by Donadi {\it et al} \cite{donadi} as the most stringent bound on $R_0$ from heating experiments reported in literature. Here, we point out that much stronger bounds on the specific heating power per unit mass have been actually reported by other authors in the context of the closely related CSL model \cite{adlervinante, carlesso2018}, implying stronger bounds on $R_0$. 

As first example, the specific power radiated by the coldest outer planets of the Solar System is indeed much lower than in neutron stars \cite{carlesso2018}. The best case is given by Neptune, for which the emission of blackbody power at equilibrium is estimated as $P \approx 20$ pW/kg, implying a lower bound $R_0 \gtrsim 3.7\times 10^{-12}$ m \cite{carlesso2018}. This value of $R_0$ is just one order of magnitude lower than the one set by x-ray emission, and almost two orders higher than the bound set by LISA Pathfinder through force noise measurements or by neutron star heating.

As a second case (more relevant to our discussion, as shown below), let us consider residual heat leak experiments performed in ultralow temperature cryostats. Adler and Vinante have derived the strongest bounds on the CSL model from heating experiments \cite{adlervinante} using a value for the unmodeled specific power $P \approx 10$ pW/kg. This figure is typically obtained in the best nuclear demagnetization cryostats, designed to reach bulk temperatures of a few $\mu$K. For concreteness, we refer to the cryostat described by Gloos {\it et al} \cite{pobell}. Here, the heat leak, and the minimum temperature achievable by the massive nuclear demagnetization stage made of copper (mass $m=17$ kg), show a well-defined time-dependent behaviour $\sim t^{-3/4}$ and $\sim t^{-3/8}$ respectively. The reduction of heat leak with time was attributed to a combination of several relaxation mechanisms, of which hydrogen ortho-para conversion is the most well-known. Small deviations from the time-dependent behaviour allow to estimate a {\it time-independent} (as expected for an hypothetical DP effect) heat leak of the order of 1 pW/mol, corresponding to about 15 pW/kg. However, note that 1 pW/mol is also the estimated contribution from cosmic muons and radioactive events, which should be therefore subtracted in order to estimate the unexplained heat leak. Taking into account the unavoidable uncertainties, we take 10 pW/kg as a reasonable estimation of the unmodeled time-independent heat leak on the copper stage. 

This value, shown as blue line in Fig. 1, leads to a bound $R_0 \gtrsim 4.6 \times 10^{-12}$ m, comparable to the one obtained from Neptune equilibrium temperature. However, the data from ultralow temperature experiments turn out to be more relevant to our discussion. The reason is that the massive copper stage of a ultralow temperature cryostat is virtually at absolute zero temperature ($T< 1$ mK) with respect to the phonon energy, meaning that the spread of the nuclear wave function is dominated by zero-point fluctuations. This is definitely not true for Neptune, indeed the temperature at the center of the planet is estimated $T \approx 5 \times 10^3$ K. For the specific experiment of Ref. \citenum{pobell}, the nuclear wave function spread can be estimated as $u_{\rm rms}=\sqrt{B/8\pi^2}=4.3 \times 10^{-12}$ m, where $B=0.146\times 10^{-20}$ m$^2$ is the Debye-Waller factor of copper at low temperature \cite{debyewaller}. $u_{\rm rms}$ is shown as a dashed line in Fig. 1. The corresponding specific power $12$ pW/kg, predicted by the DP model via Eq.~(\ref{DPheating}) by identifying $R_0$ with $u_{\rm rms}$, is of the same order of the upper limit on time-independent heat leak obtained above. 

We conclude that the hypothetical heat leak induced by the parameter-free DP model, using for $R_0$ the nuclear wave function spread \cite{penrose2014}, should be barely observable in the most accurate measurements of heat leak in ultralow temperature experiments. Given the order-of-magnitude nature the theoretical predictions and the experimental uncertainties, a conclusive exclusion will likely require an improvement of the experimental data by at least one order of magnitude. In principle, comparable or lower values of residual specific power may be measurable in similar cryostats. In particular, specifically designed experiments operated underground, and therefore insensitive to cosmic muons, should be capable of lowering significantly the upper limit on the unmodeled time-independent heat leak.

We note that bounds based on spontaneous heating effects probe the noise field at a frequency of the order of the dominant phonons \cite{adlervinante}, corresponding to the THz scale. Compared to the recent claim of exclusion of the parameter-free DP model, based on x-ray spontaneous radiation\cite{donadi}, heat leak experiments probe the DP noise field at a frequency 6 orders of magnitude lower. In this respect, they are more robust towards nonwhite modifications of the model. Indeed, to evade the bound from x-ray spontaneous emission it would be sufficient to postulate a high frequency cutoff at any frequency below $10^{18}$ Hz. Similar modifications have been often proposed in the case of the more celebrated CSL model \cite{adler}: 

A further relevant point is that, due to the very low temperature achievable by nuclear demagnetization cryostats, the bound on $R_0$ is very robust to dissipative extensions of the DP model \cite{smirne2014}, in which the noise field is associated to a dissipation and therefore can be modeled as a thermal bath with an effective temperature $T_{\rm {DP}}$. In fact, as long as one assumes astrophysically reasonable values $T_{\rm {DP}} \approx 10^0$ K, the DP noise field would effectively appear to the system as an infinite temperature bath. In this limit, the dissipative DP model provides the same predictions of the standard nondissipative version.

Finally, we note that recently proposed classical channel gravity arguments \cite{milburn} predict a similar universal spontaneous heating.  Here, the effect arises as a natural consequence in any theory in which gravity can mediate only classical information, and gravity-induced entanglement is forbidden. Such a theory would imply a deviation from the Newtonian $1/r^2$ law below a characteristic length $a$. The predicted spontaneous heating differs from the DP heating Eq.~(\ref{DPheating}) by a factor $4 \sqrt{\pi} $, with $R_0$ replaced by $a$. The experimental bound on the length parameter $a>0.9 \times 10^{-11}$ m, obtained using the data discussed in this paper, is one order of magnitude better than those inferred in Ref. \citenum{milburn}. Our work suggests a possible long-term path towards ruling out classical channel gravity models, based on ultra-accurate measurements of residual heat leak, combining ultralow temperatures, deep underground operation and very long operation time, sufficient to suppress parasitic time-dependent effects by orders of magnitude.

\begin{acknowledgments}
We would like to thank Roger Penrose for his inspiring work on the topic of wave function collapse and the role of gravity in that, which has triggered our work and that of many others. HU would like to thank Roger for scientific discussions as well as for his kind and patient advice, vision and enthusiasm for setting up a research institute on the topic. 

The authors acknowledge financial support form the EU
Horizon 2020 research and innovation program under
Grant Agreement No. 766900 [TEQ], the COST action
15220 QTSpace and the Leverhulme Trust [RPG-2016-046]. HU acknowledges support by the UK EPSRC [EP/V000624/1].
\end{acknowledgments}

\section*{Data Availability Statement}
Data sharing is not applicable to this article as no new data were created or analyzed in this study.

\bibliography{gravitycollapse}

\end{document}